\documentclass[english,10pt,final,twocolumn]{elsarticle}
\usepackage[T1]{fontenc}
\usepackage[latin9]{inputenc}
\usepackage{pdfpages}
\usepackage{amstext}
\usepackage{graphicx}

\makeatletter

\newcommand{\lyxmathsym}[1]{\ifmmode\begingroup\def\b@ld{bold}
  \text{\ifx\math@version\b@ld\bfseries\fi#1}\endgroup\else#1\fi}
\usepackage{braket}
\usepackage{mathrsfs}
\usepackage{slashed}
\usepackage{bm}
\usepackage{datetime}
\usepackage{siunitx}

\DeclareSIUnit[number-unit-product = {}]\clight{c}
\DeclareSIUnit\eVperc{\eV\per\clight}
\DeclareSIUnit\GeVpercs{\giga\eV\squared\per\clight\squared}
\DeclareSIUnit\MeVpercs{\mega\eV\per\clight\squared}
\journal{Physics Letters B}
\biboptions{numbers,sort&compress}
\usepackage[left=1.6cm,right=1.04cm,top=1.65cm,bottom=1.65cm,columnsep=25pt]{geometry}
\usepackage{lineno}
\usepackage{hyperref}  % hyper-links 
\hypersetup{breaklinks = true, colorlinks = true, allcolors = magenta}
\usepackage{setspace}
\usepackage{graphicx}
\usepackage{wrapfig}

\makeatother

\usepackage{babel}
\begin{document}

\begin{frontmatter}{}

\title{Components of polarization-transfer to a bound proton in a deuteron
\\
measured by quasi-elastic electron scattering }

\author[TAU]{D.~Izraeli\corref{cor2}}

\ead{davidizraeli@post.tau.ac.il}

\author[TAU]{I.~Yaron}

\author[Mainz]{B.S.~Schlimme}

\author[Mainz]{P.~Achenbach}

\author[Mainz]{H.~Arenh\"ovel}

\author[TAU]{A.~Ashkenazi}

\author[JSI]{J.~Beri\v{c}i\v{c}}

\author[Mainz]{R.~B\"ohm}

\author[zagreb]{D.~Bosnar}

\author[TAU]{E.O.~Cohen}

\author[Mainz]{M.O.~Distler}

\author[Mainz]{A.~Esser}

\author[zagreb]{I.~Fri\v{s}\v{c}i\'{c}\fnref{mit}}

\author[Rutgers]{R.~Gilman}

\author[TAU,nrc]{I.~Korover}

\author[TAU]{J.~Lichtenstadt}

\author[TAU,soreq]{I.~Mardor}

\author[Mainz]{H.~Merkel}

\author[Mainz]{D.G.~Middleton}

\author[JSI,Mainz]{M.~Mihovilovi\v{c} }

\author[Mainz]{U.~M\"uller}

\author[TAU]{M.~Olievenboim}

\author[TAU]{E.~Piasetzky}

\author[Mainz]{J.~Pochodzalla}

\author[huji]{G.~Ron}

\author[Mainz]{M.~Schoth}

\author[Mainz]{F.~Schulz}

\author[Mainz]{C.~Sfienti}

\author[UL,JSI]{S.~\v{S}irca}

\author[JSI]{S.~\v{S}tajner }

\author[USK]{S.~Strauch}

\author[Mainz]{M.~Thiel}

\author[Mainz]{A.~Tyukin}

\author[Mainz]{A.~Weber}

\author{for the A1 Collaboration}

\cortext[cor2]{Corresponding author}

\fntext[mit]{Present address: MIT-LNS, Cambridge, MA 02139, USA.}

\address[TAU]{School of Physics and Astronomy, Tel Aviv University, Tel Aviv 69978,
Israel.}

\address[Mainz]{Institut f\"ur Kernphysik, Johannes Gutenberg-Universit\"at, 55099
Mainz, Germany.}

\address[JSI]{Jo\v{z}ef Stefan Institute, 1000 Ljubljana, Slovenia.}

\address[zagreb]{Department of Physics, University of Zagreb, HR-10002 Zagreb, Croatia.}

\address[Rutgers]{Rutgers, The State University of New Jersey, Piscataway, NJ 08855,
USA.}

\address[nrc]{Department of Physics, NRCN, P.O. Box 9001, Beer-Sheva 84190, Israel.}

\address[soreq]{Soreq NRC, Yavne 81800, Israel.}

\address[huji]{Racah Institute of Physics, Hebrew University of Jerusalem, Jerusalem
91904, Israel.}

\address[UL]{Faculty of Mathematics and Physics, University of Ljubljana, 1000
Ljubljana, Slovenia.}

\address[USK]{University of South Carolina, Columbia, South Carolina 29208, USA.}
\begin{abstract}
We report the first measurements of the transverse ($P_{x}$ and $P_{y}$)
and longitudinal ($P_{z}$) components of the polarization transfer
to a bound proton in the deuteron via the $^{2}\mathrm{H}(\vec{e},e\lyxmathsym{\textquoteright}\vec{p})$
reaction, over a wide range of missing momentum. A precise determination
of the electron beam polarization reduces the systematic uncertainties
on the individual components to a level that enables a detailed comparison
to a state-of-the-art calculation of the deuteron using free-proton
electromagnetic form factors. We observe very good agreement between
the measured and the calculated $P_{x}/P_{z}$ ratios, but deviations
of the individual components. Our results cannot be explained by medium
modified electromagnetic form factors. They point to an incomplete
description of the nuclear reaction mechanism in the calculation.
\end{abstract}

\end{frontmatter}{}

Measurements of the polarization transfer $\vec{P}=\left(P_{x},P_{y},P_{z}\right)$
from a polarized electron to a bound nucleon by the $A(\vec{e},e\lyxmathsym{\textquoteright}\vec{p})$
reaction and their comparison to those of a free proton were suggested
as a powerful tool to observe modifications in the bound proton structure~\cite{Perdrisat}.
These require detailed calculations incorporating nuclear effects.
However, it still might be conceptually difficult to separate such
effects from internal nucleon structure changes.

The deuteron ($^{2}\mathrm{H}$) is the simplest nuclear system, with
two loosely bound nucleons. Even though it is often used as a \textquoteleft free
neutron\textquoteright{} target, measurements on the bound proton
in $^{2}\mathrm{H}$ in quasi-elastic kinematics are known to have
marked differences in comparison to those on a free proton~\cite{Gn}.
Such differences can be ascribed to final state interactions (FSI)
and other nuclear effects in the deuteron like meson exchange currents
(MEC) and isobar configurations (IC), as well as to nuclear medium
modifications of the bound proton electromagnetic form factors (FFs).
Since most of the properties of the deuteron are described very well
by calculations, it serves also as a benchmark for nuclear theory.
As such it is important to establish that present state-of-the-art
calculations indeed provide a correct description of the deuteron.
Thus, new high precision polarization transfer data can add important
information to further test and improve the calculations, since polarization
observables generally provide more sensitive tests.

Measurements of the ratio of the polarization transfer components
$P_{x}$ to $P_{z}$ ($P_{x}/P_{z}$) to a bound proton in $^{2}\mathrm{H}(\vec{e},e\lyxmathsym{\textquoteright}\vec{p})$
over a wide region of the proton missing momentum were reported in~\cite{deep2012PLB}.
Measuring the ratio, instead of the individual components, eliminates
many systematic experimental uncertainties, particularly those due
to electron beam polarization, and is sensitive to the FF ratio. The
theoretical calculation~\cite{Arenhovel} that uses free proton FFs
reproduced very well the observed deviations from the free proton
ratio, suggesting that they stem mainly from FSI. However, while the
ratio of the polarization transfer components is sensitive (almost
linearly) to the electromagnetic FFs ratio $G_{E}/G_{M}$, some nuclear
effects may cancel out in the ratio. The measured individual polarization
transfer components provide a more stringent test of the calculation.

Moreover, it was shown that the deviations of $P_{x}/P_{z}$ measured
on a bound proton in $^{2}\mathrm{H}$ from that on a free proton,
were in very good agreement with similar measurements on heavier nuclei,
when using the proton virtuality as a universal parameter for these
comparisons~\cite{ceepLet}. Such comparisons are improved by relating
the data of each nucleus to a realistic model of the deuteron data,
a process that requires the knowledge of the individual polarization
transfer components~\cite{polar2}. We introduce such a model below. 

In this work we report a new analysis of the $^{2}\mathrm{H}(\vec{e},e\lyxmathsym{\textquoteright}\vec{p})$
reaction measured at the Mainz Microtron (MAMI) over a wide range
of missing momentum of the struck proton. Results of the polarization
transfer ratio $P_{x}/P_{z}$ deduced from these measurements were
reported in~\cite{deep2012PLB}. In the new analysis, the beam polarization
that was measured periodically during the experiment was determined
in a continuous manner. The achieved accuracy was sufficient for extracting
the individual polarization transfer components $P_{x}$, $P_{y}$
and $P_{z}$, with a precision that can challenge theory. 

\begin{figure}[bh]
\begin{centering}
\includegraphics[width=0.93\columnwidth]{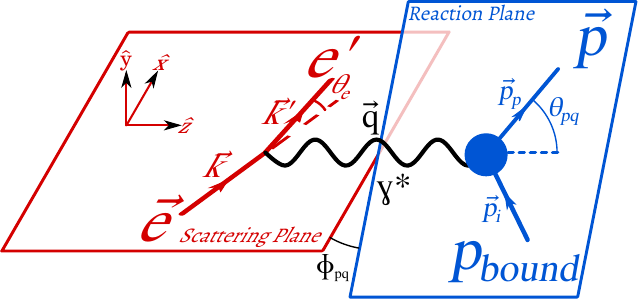}
\par\end{centering}
\caption{\label{fig:The-kinematics}The kinematics for quasi-elastic scattering
of a bound proton in a nucleus, defining the scattering and reaction
planes.}
\end{figure}

The kinematics for the quasi-free elastic scattering off a bound nucleon
is shown in Fig.~\ref{fig:The-kinematics}. The reaction plane is
determined by the momentum transfer ($\vec{q}$) and the outgoing
proton momentum ($\vec{p}_{p}$), characterized by the spherical angles
$\theta_{pq}$ and $\phi_{pq}$. The incident and scattered electron
momenta that define the scattering plane are indicated by $\vec{k}$
and $\vec{k}'$. The initial and outgoing proton momenta are indicated
by $\vec{p}_{i}$ and $\vec{p}_{p}$, respectively. The missing momentum
is $\vec{p}_{\mathrm{miss}}=\vec{q}-\vec{p}_{p}$. The missing momentum
($p_{\mathrm{miss}}$) is taken to be positive (negative) if a component
of $\vec{p}_{\mathrm{miss}}$ is parallel (anti-parallel) to the momentum-transfer
vector. In the impulse approximation with no FSI one has $\vec{p}_{\mathrm{miss}}=-\vec{p}_{i}$.
Following the convention of~\cite{deep2012PLB,ceepLet} the polarization
components reported here are perpendicular to the scattering plane
($\hat{y}$) and in the scattering plane along ($\hat{z}$) and perpendicular
($\hat{x}$) to $\vec{q}$.

The experiment was performed on the MAMI A1 beam line using a liquid
deuterium target and two of the A1 high-resolution spectrometers~\cite{a1aparatus}.
Polarized continuous wave (CW) electron beams of \SI{600}{MeV} and
\SI{630}{MeV} were used with currents of \SI{10}{\micro A}. The
target was an oblong cell (\SI{50}{mm} long, \SI{11.5}{mm} in diameter)
filled with liquid deuterium. The spectrometers have momentum acceptances
of $20-25\,\%$ with solid angles of $\SI{28}{msr}$ and were used
to detect the scattered electrons and the knocked out protons in coincidence.
The proton spectrometer was equipped with a so called ``focal-plane
polarimeter'' (FPP) placed behind its focal-plane, with a \SI{7}{cm}
thick carbon analyzer~\cite{a1aparatus,Pospischil:2000pu}. The spin
dependent scattering of the polarized proton by the carbon analyzer
enabled the determination of the transverse polarization components
at the focal plane~\cite{Pospischil:2000pu}. The polarization transfer
components at the reaction point were obtained by correcting the measured
components for the spin precession in the magnetic field of the spectrometer~\cite{Pospischil:2000pu}. 

The measurements covered two $Q^{2}$ ranges and two beam energies,
in order to span a wide range of $p_{\mathrm{miss}}$. For further
details see~\cite{supplemental}.

\begin{figure}[bh]
\includegraphics[bb=0bp 9bp 576bp 316bp,clip,width=1\columnwidth]{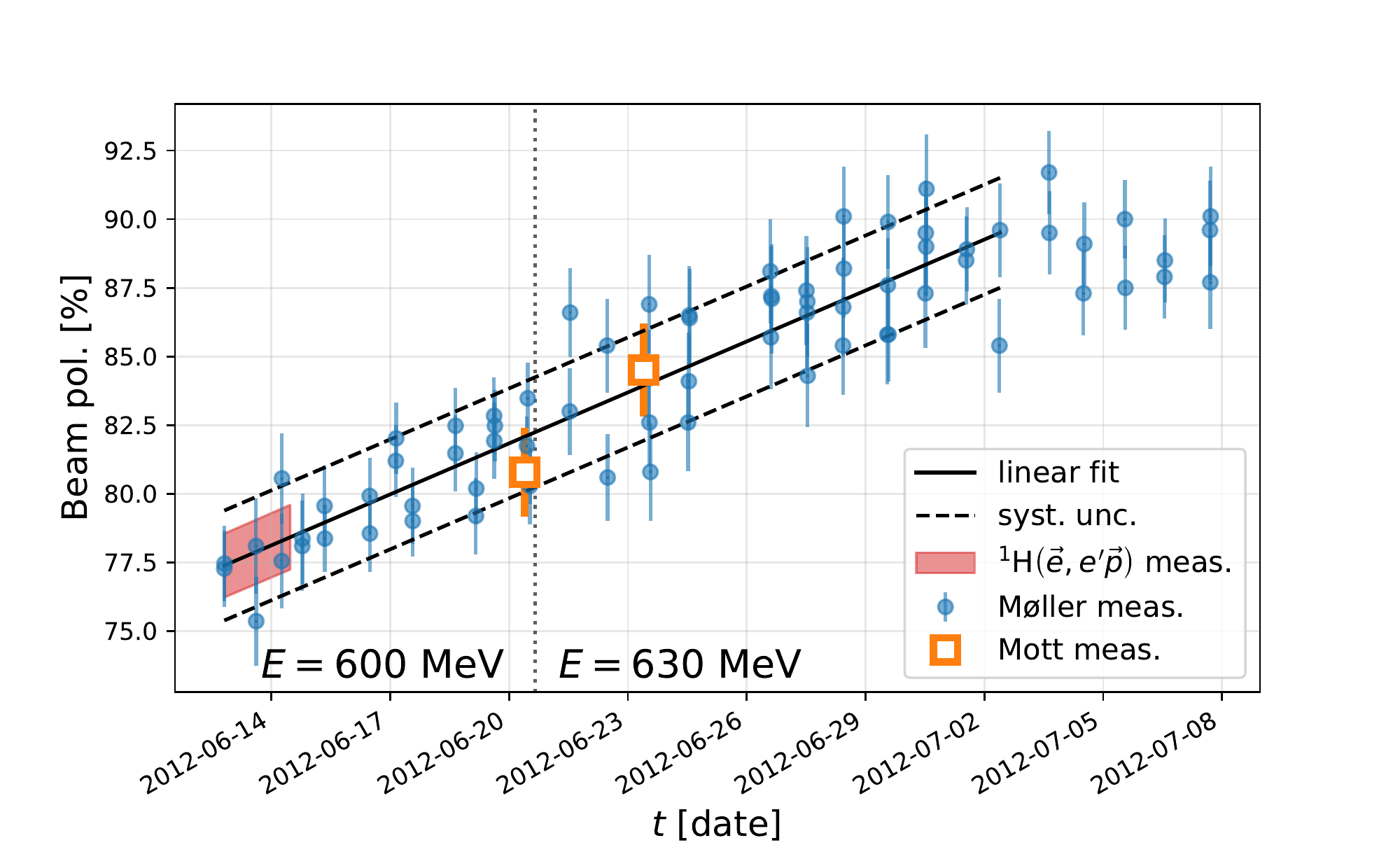}

\caption{\label{fig:The-beam-polarization}The results of the beam polarization
measurements performed with the M\o ller polarimeter. The linear
fit to the data during the data taking period and the uncertainty
band are shown as solid and dashed lines. Also shown are the Mott
and $^{1}\mathrm{H}(\vec{e},e\lyxmathsym{\textquoteright}\vec{p})$
measurements. See text for details. }
\end{figure}

The beam polarization at MAMI is obtained by using strained GaAs photocathodes.
The beam polarization increases throughout the usage of the single
cathode, due to the decrease of its quantum efficiency~\cite{MAMI_NIM_1997,ALLEY1995}.
The beam polarization was measured daily using a M\o ller polarimeter
located upstream of the target cell. The measurements during the three
weeks of the experimental run are shown in Fig.~\ref{fig:The-beam-polarization},
where the slow increase in time is clearly seen. The statistical uncertainty
on each measurement is about $1.5\%$, but there is an overall systematic
uncertainty of a few percent, due to the calibration of the detectors
in the M\o ller polarimeter. These measurements were fitted by a
linear function ($\chi^{2}/\mathrm{ndf}=60.2/61$). 

In addition to the M\o ller measurements, the beam polarization was
measured for each beam energy by Mott scattering. The Mott and M\o ller
measurements are consistent (see Fig.~\ref{fig:The-beam-polarization}).

To determine the overall normalization of the beam polarization and
the analyzing power, we use measurements of the polarization transfer
to a free proton by $^{1}\mathrm{H}(\vec{e},e\lyxmathsym{\textquoteright}\vec{p})$,
which were performed in the beginning of the run (indicated by the
red time interval in Fig.~\ref{fig:The-beam-polarization}) at $Q^{2}=\SI{0.4}{\GeVpercs}$.
Assuming the FF ratio $G_{E}/G_{M}$ for a free proton from the parameterization
of~\cite{Bernauer} (estimated uncertainty of $0.1\%$), we deduced
the beam polarization from the polarization transfer for these measurements.
The resultant overall normalization was $1.00\pm0.01$. The uncertainty
is dominated by the statistical uncertainty of our hydrogen measurements.
The band in Fig.~\ref{fig:The-beam-polarization} shows the overall
uncertainty of the beam polarization.

In the analysis of the polarization components, cuts were applied
to identify coincident electrons and protons that originate from the
deuterium target, and to ensure good reconstruction of tracks in the
spectrometers and the FPP. Only events that scatter by more than $\ang{10}$
in the FPP were selected (to remove Coulomb scattering events in the
FPP carbon analyzer). For each event we used the beam polarization
obtained from the time-dependence fit. 

Time-independent corrections to the polarization transfer measurements
(acceptance, detector efficiency, target density, etc.) are largely
canceled out by the frequent flips of beam helicity. Contributions
to the systematic uncertainty due to the carbon analyzing power and
FPP efficiency are well below the statistical uncertainty. The total
systematic uncertainties in $P_{x}$ and $P_{z}$ are estimated to
be about $2\%$ and are due to beam polarization uncertainty (canceled
in the $P_{x}/P_{z}$ ratio) and the reaction vertex reconstruction
(which dominates both the momentum resolutions and the spin-precession
evaluation). The systematic uncertainty on $P_{y}$ is estimated to
be comparable to the statistical one.

In our previous analysis~\cite{deep2012PLB}, where $(P_{x}/P_{z})^{^{2}\mathrm{H}}/(P_{x}/P_{z})^{^{1}\mathrm{H}}$
was compared to $(P_{x}/P_{z})^{A}/(P_{x}/P_{z})^{^{1}\mathrm{H}}$
in other nuclei, it was shown that the nucleon virtuality, defined
as\begin{linenomath}
\begin{equation}
\nu\equiv\Big(m_{d}c-\sqrt{m_{n}^{2}c^{2}+p_{\text{miss}}^{2}}\Big)^{2}-p_{\text{miss}}^{2}-m_{p}^{2}c^{2},
\end{equation}
\end{linenomath} can serve as a universal parameter for such comparisons.
The ratio of the measurements to those of a free proton eliminates
some of the variations that are due to kinematics and $Q^{2}$ dependence.
Using $\nu$ as a parameter was shown to give a good universal description
of the data, which was further established in new measurements on
$^{12}\mathrm{C}$~\cite{ceepLet}.

\begin{figure}[th]
\includegraphics[bb=0bp 9bp 471bp 520bp,clip,width=1\columnwidth]{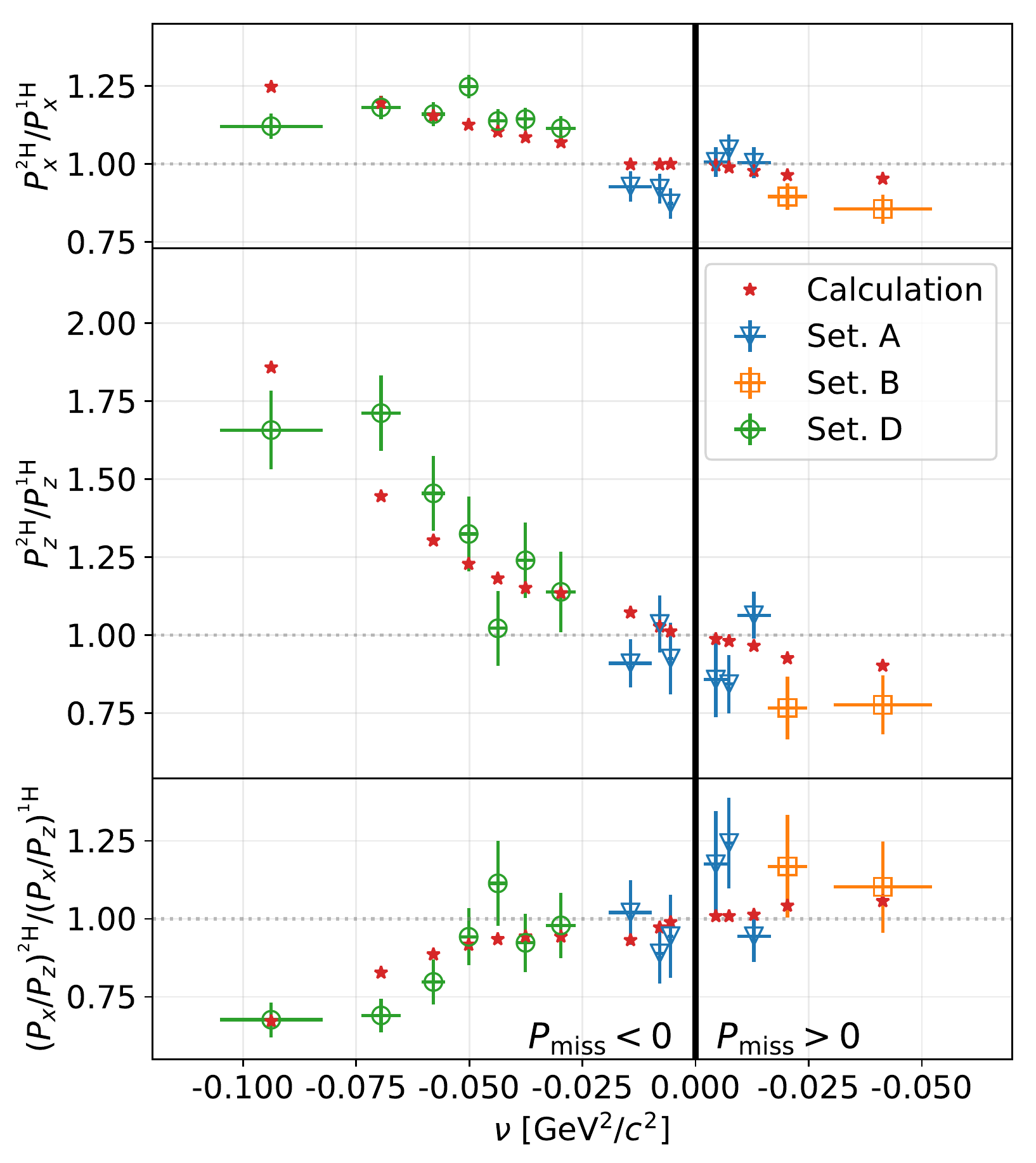}

\caption{\label{fig:The-measured-ratios}The measured ratios $P_{x}^{^{2}\mathrm{H}}/P_{x}^{^{1}\mathrm{H}}$,
$P_{z}^{^{2}\mathrm{H}}/P_{z}^{^{1}\mathrm{H}}$ and the double-ratio,
$(P_{x}/P_{z})^{^{2}\mathrm{H}}/(P_{x}/P_{z})^{^{1}\mathrm{H}}$,
as a function of the proton virtuality, $\nu$. The virtuality dependence
is shown separately for positive and negative missing momenta. The
symbols for the data of this work correspond to the different kinematical
settings, see~\cite{supplemental}. Also shown is a calculation for
the deuteron~\cite{Arenhovel} (see text for details).}
\end{figure}

We follow this work, and present the ratio of the polarization components
$P_{x}$ and $P_{z}$ (and their ratio $P_{x}/P_{z}$) to the corresponding
values of a free proton as function of $\nu$, in Fig.~\ref{fig:The-measured-ratios}.
The proton values were obtained from the global parameterization of
Bernauer \emph{et al.}~\cite{Bernauer}. In the analysis the ratio
was taken event by event, and averaged over the virtuality bin. The
data in Fig.~\ref{fig:The-measured-ratios} are shown separately
for positive and negative missing momenta to show possible differences
predicted by the calculation.

In this analysis, the $P_{y}$ component of the polarization-transfer,
which vanishes for the free proton, was also determined, resulting
in very small values. Its consideration is essential for proper determination
of $P_{x}$ and $P_{z}$. Since it cannot be compared to the free
proton it is not shown in Fig.~\ref{fig:The-measured-ratios}. Also
shown in Fig.~\ref{fig:The-measured-ratios} are the calculated ratios~\cite{Arenhovel}.
The calculation uses free proton FFs, and was carried out over the
entire data for each event, and averaged over the corresponding bins.

The data indicate that the deviation of $(P_{x}/P_{z})^{^{2}\mathrm{H}}/(P_{x}/P_{z})^{^{1}\mathrm{H}}$
from unity is mainly due to the $P_{z}$ component, which according
to the calculation, is more sensitive to FSI and relativistic corrections.
The calculation is in very good overall agreement with the data. Since
medium modifications in the bound proton structure are inferred from
deviations of measured data from calculations, a careful comparison
between the data and the calculation is necessary.
\begin{figure}[th]
\includegraphics[bb=0bp 8bp 471bp 677bp,clip,width=1\columnwidth]{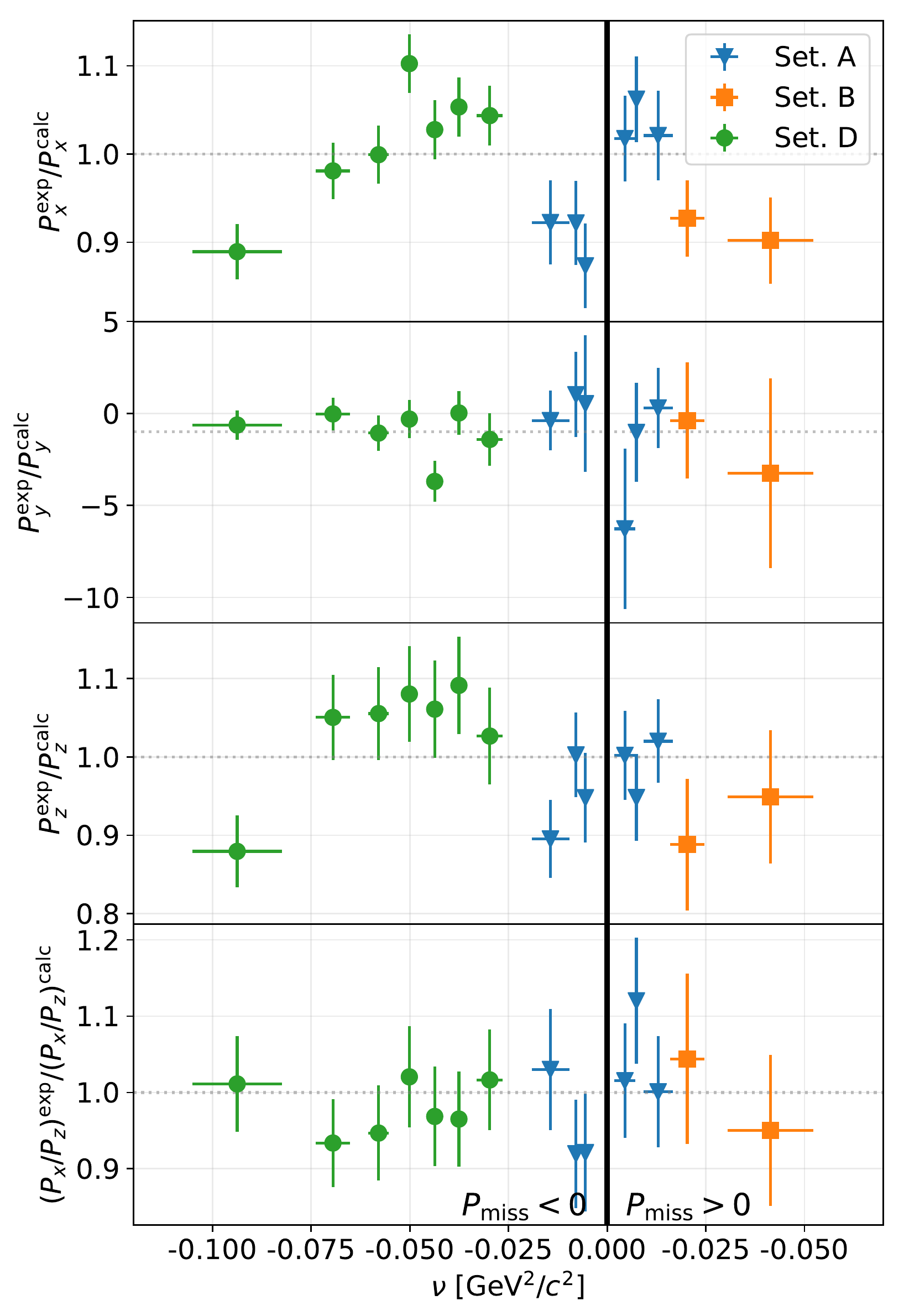}

\caption{\label{fig:The-ratios}The ratio of the measured to calculated~\cite{Arenhovel}
polarization transfer ratios for the components: $P_{x}$, $P_{y}$,
$P_{z}$ and the ratio $P_{x}/P_{z}$ (top to bottom panels, respectively).
We use the same symbols as in Fig.~\ref{fig:The-measured-ratios}.}
\end{figure}

The comparison was performed by investigating the ratio of the measurements
to the calculation event-by-event over the entire data set. These
ratios are shown in Fig.~\ref{fig:The-ratios}. The measured $P_{x}/P_{z}$
ratio is in very good agreement with the calculation ($p_{\mathrm{value}}=0.91$).
The deviations of about $10\%$ observed in~\cite{deep2012PLB} are
reduced to less than $1\%$. This is due to: (a)~a comparison of
the data with the calculation~\cite{Arenhovel} in the lab frame\footnote{We note that in~\cite{deep2012PLB} the data (in the laboratory frame)
were compared to the calculation in the c.m.\ frame resulting in
a 10\% discrepancy. A correct comparison will reduce the deviation
to about 5\%.}, (b)~improvement of the analysis by using the time-dependent polarization
to weigh each event, and (c)~a new procedure that compares all three
components with the calculation that results in a better overall fit.
As can be seen in Fig.~\ref{fig:The-ratios}, we observe differences
between the data and the calculation of the $P_{x}$ and $P_{z}$
values. Free proton $P_{x}$ and $P_{z}$ are functions of the FF
ratio $G_{E}/G_{M}$, which become linear for the $P_{x}/P_{z}$ ratio.
Similarly, for the deuteron, the calculation that uses free nucleon
FFs exhibits the same behavior. The good agreement between the measured
and the calculated $P_{x}/P_{z}$ ratios indicates no need for modifications
in $G_{E}/G_{M}$. This does not exclude modifications of the individual
FFs which keep the ratio intact. Excluding FF modifications, the deviations
of the measured individual components from the calculated ones, suggest
that the nuclear effects and/or relativistic corrections included
in the calculation should be improved.
\begin{figure}[th]
\includegraphics[bb=0bp 8bp 471bp 520bp,width=1\columnwidth]{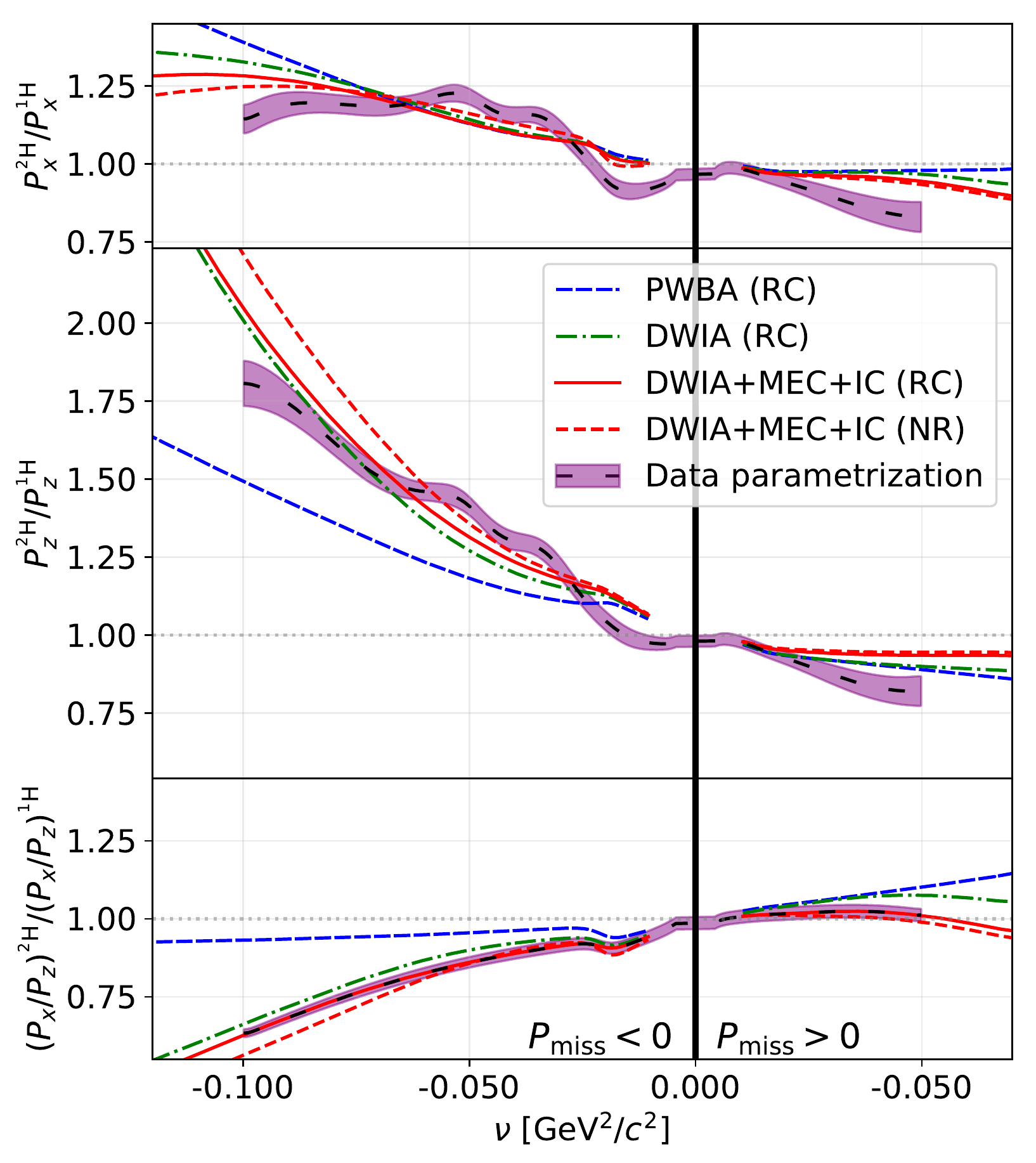}

\caption{\label{fig:Calculations-of-HA}Calculations of~\cite{Arenhovel}
with FSI (DWIA) and without (PWBA), with first order relativistic
corrections (RC), and without (NR). Also shown is the effect of adding
MEC and IC corrections. The data parametrization, obtained by using
multiplication factors that adjust the calculation to the measurement,
is shown as a dashed line with an uncertainty band of one standard
deviation.}
\end{figure}

Comparisons of polarization transfer data amongst several nuclei are
improved by relating the data of each nucleus to a realistic model
of the deuteron data, a process that requires the knowledge of the
individual polarization transfer components~\cite{polar2}. The accuracy
of this process is limited if one averages the results within finite
size bins. Using a continuous parameterization of the $^{2}\mathrm{H}$
data was found to have significant advantages and to be very useful
for such a comparison with the $^{12}\mathrm{C}$ data~\cite{ceepLet}.
We introduce in Fig.~\ref{fig:Calculations-of-HA} a parameterization
of the data as a function of the nucleon virtuality. The number of
fitted parameters was optimized to avoid over-fitting. Shown are the
ratios of the individual components to those of the free proton, as
well as the double ratio $(P_{x}/P_{z})^{^{2}\mathrm{H}}/(P_{x}/P_{z})^{^{1}\mathrm{H}}$,
which were obtained by using multiplication factors to adjust the
calculation to the data on an event by event basis. Also shown are
the calculations with different ingredients as well as the full calculation~\cite{Arenhovel},
from which one can infer that the main deviation from the free proton
is due to FSI, as can be seen by comparing the PWBA and DWIA curves.
The full parameterization of the data is available in~\cite{supplemental}.

To summarize, we combined our beam polarization measurements with
hydrogen measurements to reduce the uncertainty on the beam polarization.
This enables the determination of the individual components $P_{x}$,
$P_{y}$ and $P_{z}$ of the polarization transfer with very good
precision. The deviations of the measured $P_{x}$ and $P_{z}$ from
the corresponding values for a free proton show that the observed
deviation of the $P_{x}/P_{z}$ ratio is mainly due to $P_{z}$, which
seems to be more sensitive to FSI and relativistic corrections than
$P_{x}$.

The comparison of our data to the calculation shows very good agreement
for the $P_{x}/P_{z}$ ratio, but discrepancies in the individual
components. These observations rule out FF changes in the calculation
(except for modifications which keep the ratio intact) and suggest
the need of improved calculations of FSI and relativistic corrections
in the deuteron. 

While the data cover a relatively wide range in virtuality (and $p_{\mathrm{miss}}$),
further extension of the kinematic range, particularly in the positive
$p_{\mathrm{miss}}$ region, may further challenge calculations, and
are thus important to our understanding of the nuclear mechanisms
at work. 

We would like to thank the Mainz Microtron operators and technical
staff for the excellent operation of the accelerator. This work is
supported by the Israel Science Foundation (Grant 390/15) of the Israel
Academy of Arts and Sciences, by the Israel Ministry of Science, Technology
and Space, by the Deutsche Forschungsgemeinschaft (Collaborative Research
Center 1044), by the Slovenian Research Agency (research core funding
No.~P1\textendash 0102), by the U.S. National Science Foundation
(PHY-1505615), and by the Croatian Science Foundation Project No.~1680.
\newpage{}\bibliographystyle{elsarticle-num}
\addcontentsline{toc}{section}{\refname}\bibliography{comp}

\clearpage{}

\newpage{}\newpage{}\newpage{}

\includepdf[pages=-]{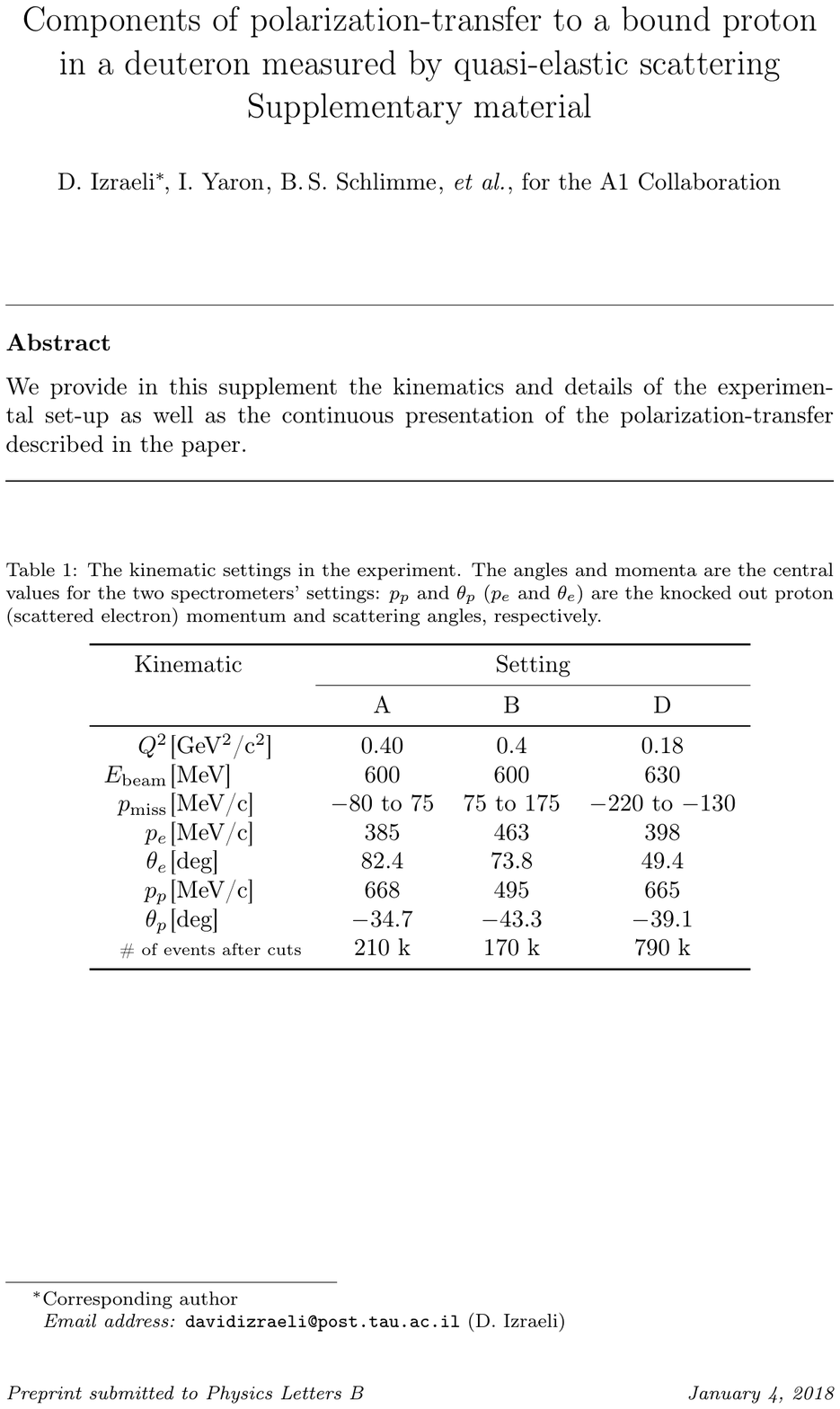}
\end{document}